# The free states-related Fermi pocket of cuprate superconductors

Tian De Cao[*]

*Department of Physics, Nanjing University of Information Science & Technology, Nanjing 210044, China*

**Abstract**

To stress the effect of the pairing position deviating from the Fermi level, we must investigate the pairs in the wave vector space, and then we use the dynamic equation to study some correlation functions. This article shows that the Fermi pocket is related to the effect of free electron states on the ARPES experiment. This also leads us to understand why the Fermi arc appears in Bi2212 while the Fermi pocket appears in Bi2201 with the valence bandwidth and the work function known for them.

## Ⅰ. Introduction

The Fermi pocket is suggested by the d-density wave [1,2] or other order-based mechanisms, and it is observed in the underdoped Bi2201[3]. However, there are some deviation between the theory prediction and the experiment result. Moreover, the Fermi arc is discovered in the underdoped Bi2212 [4,5]. Why does the Fermi pocket appear in the underdoped Bi2201 not in the Bi2212? In the previous paper [6], it is shown that the electronic structures of cuprate superconductors favor the pairs-based pseudogap appearing in the antinodal region, and the relation between superconductivity and pseudogap can be understood well, despite some physicists suggested the order-based pseudogap [7,8]. On the basis of same ideas, this article shows that the Fermi arc and pocket can be understood.

## Ⅱ. Review of ARPES principle

The quantum measured by the ARPES experiment is the hopping probability from a localized orbital to a free state, while both the localized orbital and the free state all belong to the same Hamiltonian $H_0$ as shown in Eq.(2). The total Hamiltonian is $H = H_0 + H'$, $H'$ describes the interaction between electrons and photons. Because the photons can increase the probability of the electrons occupying free-particle states, when we intend to discuss the electron states which are not too lower than the Fermi energy level, we should also consider the free electron states, and we derived the formula of photoemission intensity [9]

$$I(\vec{k}, E_k^{kin}) = C' M_{int}^2 k \, n_F^a(E_k^{kin} + \phi - h\nu) \cdot A_a(\vec{k}, \sigma, E_k^{kin} + \phi - h\nu) \tag{1}$$

where $E_{kin}$ is the kinetic energy of the photoelectron, $h\nu$ is the photon energy, and $\phi$ is the work function. It is necessary to note that $A_a(\vec{k}, \sigma, \omega)$ is the one-particle spectral function for the electron systems both free states and localized electron orbitals with the Hamiltonian $H_0$ as discussed in the previous literature [10](the so-called "state outside a solid" in the previous paper seems vague, because the free state is also the state inside a solid.), and the energy conservation law $E_k^{kin} + \phi - h\nu - E_B^{\vec{k}} = 0$ is used, where $E_B^{\vec{k}}$ are the energies of electrons ($E_B^{\vec{k}} = 0$ at $E_F$) inside the solid.

*Corresponding author.

*E-mail address: tdcao@nuist.edu.cn (T. D. Cao).

*Tel: 011+86-13851628895

To understand the spectral function $A_a$, we take the simple Hamiltonian

$$H_0 = \sum_{\vec{k},\sigma} \omega_k c_{\vec{k}\sigma}^+ c_{\vec{k}\sigma} + \sum_{\vec{k},\sigma}(M_{\vec{k}} b_{\vec{k}\sigma}^+ c_{\vec{k}\sigma} + h.c.) + \sum_{\vec{k},\sigma} \xi_{\vec{k}} b_{\vec{k}\sigma}^+ b_{\vec{k}\sigma} + H_{b-b} \tag{2}$$

where $\omega_k = E_k^{kin} + \phi - \mu$, $\xi_{\vec{k}} = \varepsilon_{\vec{k}} - \mu$, $b_{\vec{q}\sigma}$ destroy an electron in $\vec{q}$ state of spin $\sigma$ inside the solid, $c_{\vec{k}\sigma}$ destroy an electron in the free states, and $H_{b-b}$ represent the interactions between electrons inside the solid. Eq.(2) requires $k_\perp = 0$, while we extend it to $k_\perp \ll k$, and this is met in some experiments. The b-band (it does not strictly correspond to the Cud or Op states) in Eq.(2) expresses the energy band near the Fermi energy in the $CuO_2$ planes of the p-type cuprate superconductors. Although the properties of the $CuO_2$ planes may be described by a multi-band model, the electronic feature near the Fermi surface could be described affectively by the a-model.

## III. Calculation

To find the obvious effect of the overlap matrix element $M_{\vec{k}}$ (another element, $M_{int}$, is not discussed in this article) in the pseudogap phase, we take the model

$$H_0 = \sum_{k,\sigma} \omega_k c_{k\sigma}^+ c_{k\sigma} + \sum_{k,\sigma}(M_k b_{k\sigma}^+ c_{k\sigma} + h.c.) + \sum_{k,\sigma} \xi_k b_{k\sigma}^+ b_{k\sigma} + U\sum_l b_{l\uparrow}^+ b_{l\uparrow} b_{l\downarrow}^+ b_{l\downarrow} \quad (3)$$

where we have denoted wave vector $\vec{k}$ as $k$, $k \equiv \vec{k}$. To compare with experiments in detail, this model should be improved. However, we are only interested in the qualitative relation of some quanta. If we introduce the charge operator $\hat{\rho}(q) = \frac{1}{2}\sum_{k,\sigma} b_{k+q\sigma}^+ b_{k\sigma}$ and the spin operator $\hat{S}(q) = \frac{1}{2}\sum_{k,\sigma} \sigma b_{k+q\sigma}^+ b_{k\sigma}$, Eq. (3) is rewritten in the form

$$H_0 = \sum_{k,\sigma} \omega_k c_{k\sigma}^+ c_{k\sigma} + \sum_{k,\sigma}(M_k b_{k\sigma}^+ c_{k\sigma} + h.c.) + \sum_{k,\sigma} \xi_k b_{k\sigma}^+ b_{k\sigma} + \sum_q U\hat{\rho}(q)\hat{\rho}(-q) - \sum_q U\hat{S}_z(q)\hat{S}_z(-q) \quad (4)$$

To discuss the effect of $M_k$, we define the functions

$$G_b(k,\sigma,\tau-\tau') = -<T_\tau b_{k\sigma}(\tau) b_{k\sigma}^+(\tau')>$$

$$F^+(k,\sigma,\tau-\tau') = <T_\tau b_{k\sigma}^+(\tau) b_{\bar{k}\bar{\sigma}}^+(\tau')>$$

$$F(k,\sigma,\tau-\tau') = <T_\tau b_{\bar{k}\bar{\sigma}}(\tau) b_{k\sigma}(\tau')>$$

The model (4) is similar to the Hubbard model, but it includes the free states which have energy levels higher above the chemical potential. The Hubbard model has not superconductivity on the basis of the evidence of off-diagonal long range order (ODLRO) [11], while other calculations found superconductivity. Moreover, we suggest that the superconductivity is due to the pairs on the Fermi surface and the ODLRO evidence is not reliable.

The doubly occupied states should be neglected for $U \to \infty$. However, we are interested in the states near the Fermi surface, thus $U$ is not too large for the doped cuprate superconductors, we establish their dynamic equations and find the two-particle Green's functions such as $<T_\tau \hat{S}(q) b_{k+q\sigma} b_{k\sigma}^+(\tau')>$ and $<T_\tau \hat{\rho}(q) b_{k+q\sigma} b_{k\sigma}^+(\tau')>$ in these equations. Moreover, we proceed to establish the dynamic equations of $<T_\tau \hat{S}(q) b_{k+q\sigma} b_{k\sigma}^+(\tau')>$ and $<T_\tau \hat{\rho}(q) b_{k+q\sigma} b_{k\sigma}^+(\tau')>$ and use the cut-off approximation. This is not the Hertree-Fock approximation because we consider the effect of correlations, and we obtain

$$[-i\omega_n + \tilde{\xi}_k + \frac{|M_k|^2}{i\omega_n - \omega_k} + \sum_q \frac{P(k,q,\sigma)}{i\omega_n - \xi_{k+q}}]G_b(k\sigma,i\omega_n)$$

$$= -1 + \frac{V(0)<\hat{\rho}(0)>}{-i\omega_n + \xi_k} + \sum_q \frac{\xi_{k+q} - \xi_k}{-i\omega_n + \xi_{k+q}} UF(k+q\sigma,\tau=0)F^+(\bar{k}\bar{\sigma},i\omega_n) \quad (5)$$

$$[-i\omega_n - \tilde{\xi}_k + \sum_q \frac{P(k,q,\sigma)}{i\omega_n + \xi_{k+q}}]F^+(k\sigma,i\omega_n) = \sum_q \frac{\xi_{k+q} - \xi_k}{-i\omega_n - \xi_{k+q}} UF^+(k-q,\sigma,\tau=0)G_b(\bar{k}\bar{\sigma},i\omega_n) \quad (6)$$

To arrive at the Eq. (5) and (6), we introduced theses functions

$$\tilde{\xi}_k = \xi_k + \frac{1}{2}Un_b$$

$$R_{(\pm)}(k,\sigma,i\omega_n) = \sum_q \frac{P(k,q,\sigma)}{i\omega_n \pm \xi_{k+q}}$$

$$\Delta_{(\pm)}(k,\sigma,i\omega_n) = \sum_q \frac{\xi_{k+q} - \xi_k}{-i\omega_n \pm \xi_{k+q}} UF(k+q\sigma,\tau=0)$$

$$\Delta^+_{(\pm)}(k,\sigma,i\omega_n) = \sum_q \frac{\xi_{k+q} - \xi_k}{-i\omega_n \pm \xi_{k+q}} UF^+(k+q\sigma,\tau=0)$$

$$P(k,q,\sigma) = U^2 <\hat{S}(-q)\hat{S}(q)> + U^2 <\hat{\rho}(-q)\hat{\rho}(q)> - 2\sigma U^2 <\hat{\rho}(-q)\hat{S}(q)>$$

where $n_b$ is the b-electron number at each site, $<\hat{S}(-q)\hat{S}(q)> \equiv <T_\tau \hat{S}(-q,\tau)\hat{S}(q,\tau-0)>$ and so on. It is shown that the function $G$, $F$ and $F^+$ depend on the spin index $\sigma$ due to the spin-charge correlation function $<\hat{\rho}\hat{S}>$, while the spin dependence does not affect our main result for non-magnetic materials, and it will be neglected below. Thus we find

$$\{i\omega_n - \tilde{\xi}_k - \frac{|M_k|^2}{i\omega_n - \omega_k} - R_{(-)}(k,i\omega_n) - \Delta^+_{(-)}(k,i\omega_n)\Delta_{(+)}(k,i\omega_n)/[i\omega_n + \tilde{\xi}_k - R_{(+)}(k,i\omega_n)]\}G_b(k,i\omega_n)$$

$$= 1 + \frac{n_b U/2}{i\omega_n - \xi_k} \quad (7)$$

$$\{[i\omega_n + \tilde{\xi}_k - R_{(+)}(k,i\omega_n)][i\omega_n - \tilde{\xi}_k - \frac{|M_k|^2}{i\omega_n - \omega_k} - R_{(-)}(k,i\omega_n)] - \Delta^+_{(-)}(k,i\omega_n)\Delta_{(+)}(k,i\omega_n)\}F^+(k,i\omega_n)$$

$$= -\Delta^+_{(-)}(k,i\omega_n)[1 + \frac{n_b U/2}{i\omega_n - \xi_k}] \quad (8)$$

A detail analysis shows that a non-zero pairing temperature $T_{pair} > 0K$ can be found with Eq. (8), and the highest pairing temperature will appear when the gap function $\Delta^+_{(-)}\Delta_{(+)}$ have the d-wave or the similar non-s wave symmetry. A detail discussion can be found in the previous paper[6] if we note that the pairing temperature is dominated by $M_k = 0$. Therefore, to discuss the electronic structure near the node, we assume $\Delta^+_{(-)}\Delta_{(+)} = 0$ in the node ($k_x = \pm k_y$) even if for $T < T_{pair}$, this electronic structure can be extended to the small value of $\Delta^{+}_{(-)}\Delta_{(+)}$ in the node, and we arrive at

$$G_b(k,i\omega_n) = [1 + \frac{n_b U/2}{i\omega_n - \xi_k}]/[i\omega_n - \tilde{\xi}_k - \frac{|M_k|^2}{i\omega_n - \omega_k} - R_{(-)}(k,i\omega_n)] \tag{9}$$

The form of $G_b$ in Eq.(9) shows that the excitation energies near (or on) the Fermi surface are determined by

$$E_k - \tilde{\xi}_k - \frac{|M_k|^2}{E_k - \omega_k} - R_{(-)}(k,E_k) = 0 \tag{10}$$

The Eq.(10) can be solved by successive iteration. Let $E_k^{(0)} - \tilde{\xi}_k - |M_k|^2/(E_k^{(0)} - \omega_k) = 0$, we get $E_{k,(\pm)}^{(0)} = \frac{1}{2}[\omega_k + \tilde{\xi}_k \pm \sqrt{(\omega_k - \tilde{\xi}_k)^2 + |M_k|^2}]$, thus $E_{k,(\pm)}^{(n)} = \tilde{\xi}_k + \frac{|M_k|^2}{E_{k,(\pm)}^{(n-1)} - \omega_k} + R_{(-)}(k,E_{k,(\pm)}^{(n-1)})$. We find $E_{k_1,(+)}^{(n)} - E_{k_2,(-)}^{(n)} = 0$ if $E_{k_1,(+)}^{(0)} - E_{k_2,(-)}^{(0)} = 0$, thus we will pay close attention to the zero order approximation $E_{k,(\pm)}^{(0)}$. To clearly see the related result, one can first take $M_k = 0$ and find $E_{k_1,(+)}^{(0)} - E_{k_2,(-)}^{(0)} = \omega_{k_1} - \tilde{\xi}_{k_2} = E_{k_1}^{kin} + \phi - \varepsilon_{k_2}$. Because the width of $\varepsilon_{k_2}$ is $2t_p$ (the bandwidth of the electron energy band in CuO$_2$ planes of cuprate superconductors), thus it could be $E_{k_1,(+)}^{(0)} - E_{k_2,(-)}^{(0)} = 0$ for some wave vectors when $\phi \leq 2t_p$. Therefore, $E_{k,(+)}^{(0)} = 0$ for $k = k_1$, $k_2$, $k_3$ ...if $E_{k,(-)}^{(0)} = 0$ for $k = k_1'$, $k_2'$, $k_3'$ ...around the node, and this case will lead to the Fermi pocket. However, $E_{k_1,(+)}^{(0)} - E_{k_2,(-)}^{(0)} > 0$ for $\phi \gg 2t_p$, and this case will lead to the Fermi arc.

## IV. Explanation of experiment results

The Fermi arc is an open-ended gapless section in the large Fermi surface, while the pocket shows the small closed Fermi surface features. We arrive at these results:

The Fermi arc can be understood. As discussed above, $E_{k_1,(+)}^{(0)} - E_{k_2,(-)}^{(0)} > 0$ for $\phi > 2t_b$, this means $E_{k,(+)}^{(n)} > 0$ for any wave vector if $E_{k,(-)}^{(n)} = 0$ for $k = k_1'$, $k_2'$, $k_3'$ ...around the node, thus only one Fermi segment could be found in each nodal region. It should be noted that the energy $E_k = 0$ means $k = k_F$. For the p-type cuprate superconductors, we can explain $b_{\bar{q}\sigma}$ as the operator which destroys an hole in the CuO$_2$ planes, thus a hole Fermi arc could be observed in the hole doped cuprates (for the appropriate structure of $\varepsilon_{\bar{k}}$). The valence bandwidth of Bi2212 is estimated to be 1.5 eV, the work function of Bi2212 is estimated to be 4.3eV[10], and the Fermi arc feature of Bi2212 has been observed [12] (their arguments also support our results). These consistencies in the Fermi arc, the bandwidth and the work function support our results above.

The Fermi pocket in the nodal region could be explained, too. As discussed above, it could be $E_{k,(+)}^{(0)}=0$ and $E_{k,(-)}^{(0)}=0$, this results in $E_{k,(+)}^{(n)}=0$ for $k=k_1$, $k_2$, $k_3$ … when $E_{k,(-)}^{(n)}=0$ for $k=k_1^{'}$, $k_2^{'}$, $k_3^{'}$ …in the nodal region, and this gives two Fermi segments at each nodal region. That the two Fermi segments form a close pocket is because $k_N = k_N^{'}$ for few wave vectors if the pairs are far from the nodal region. The bandwidth of Bi2201 is estimated to be 6.5 eV[13], the work function of Bi2212 is estimated to be 4.1eV [14], and the Fermi pocket of Bi2201 is observed in ARPES, these consistencies in ARPES measurements also support our results.

The Eq.(10) may have three solutions, and the coexistence of Fermi arc and Fermi pocket may be understood. However, this case is so complex that we will not discuss it in detail.

## Ⅴ. Conclusion

In the discussion above, our results are based on a basic suggestion that the pseudogap appears in the antinodal region. This idea is also consistent with other experiments as discussed in our previous articles [6]. Although the Hubbard-like Hamiltonian has been examined many times in both pseudogap and superconductivity, our work is based on a different idea for the moderate on-site interaction.

In summary, the Fermi arc, the Fermi pocket, and the pseudogap can be explained consistently with the electronic structures of superconductors. Moreover, this article supports further that whether the pairs lead to superconductivity or pseudogap is determined by the pairing position.